\documentclass[preprint]{emulateapj}

\newcommand{\HI}{H \textsc{i} }
\newcommand{\HII}{H \textsc{ii} }

\usepackage{apjfonts}

\begin{document}

\title{The Structure and Metallicity Gradient in the Extreme Outer
Disk of NGC~7793}

\shorttitle{The Outer Disk of NGC~7793}
\shortauthors{Vlaji\'c, Bland-Hawthorn \& Freeman}

\author{M. Vlaji\'c}

\affil{Astrophysikalisches Institut Potsdam, An der Sterwarte 16, D-14482 Potsdam, Germany}
\affil{Sydney Institute for Astronomy, School of Physics, University
of Sydney, NSW 2006, Australia}
\email{vlajic@aip.de}

\author{J. Bland-Hawthorn}
\affil{Sydney Institute for Astronomy, School of Physics, University
of Sydney, NSW 2006, Australia}

\and

\author{K. C. Freeman}
\affil{Mount Stromlo Observatory, Private Bag, Woden, ACT 2611,
Australia}

\begin{abstract}

Studies of outer regions of spirals disks are fundamental to our
understanding of both the process of galaxy assembly and the
subsequent secular evolution of galaxies.  In an earlier series of
papers we explored the extent and abundance gradient in the outer disk
of NGC~300 and found an extended purely exponential disk with a
metallicity gradient which flattens off in the outermost regions.  We
now continue the study of outskirts of pure disk spirals with another
Sculptor Group spiral, NGC~7793.  Using Gemini Multi Object
Spectrograph camera at Gemini South, we trace the disk of NGC~7793
with star counts out to $\sim9$ scale lengths, corresponding to 11.5
kpc at our calibrated distance of $3.61\pm0.53$ Mpc.  The outer disk
of NGC~7793 shows no evidence of a break in its light profile down to
an effective surface brightness of $\sim30$ mag arcsec$^{-2}$ ($\sim3$
mag arcsec$^{-2}$ deeper than what has been achieved with surface
photometry) and exhibits a non-negative abundance gradient within the
radial extent of our data.

\end{abstract}


\section{INTRODUCTION}
\label{sec:introduction}

The study of the extreme outer disks of nearby spirals has produced
some remarkable findings in recent years.  These advances have been
brought together by a confluence of observational and theoretical
discoveries.  On the observational side, star counts offer a superior
method for tracing faint stellar populations in the outskirts of
galaxies, compared to traditional surface photometry
\citep{pritchet94,ferguson02,blandhawthorn05,irwin05,dejong07,radburnsmith11a}.
While surface photometry is limited by diffuse background sources at
the level of $\sim27$ mag arcsec$^{-2}$, observations of resolved
stellar populations do not suffer from the same limitations and allow
us to reach $3-5$ mag arcsec$^{-2}$ deeper in effective surface
brightness.  On the other hand, the N-body/SPH and semi-analytical
models of disk formation are now able to address the importance of
secular effects in the evolution of spirals.  Scattering of stars by
transient spiral arms first studied by \citet{sellwoodbinney02} could
potentially have a significant effect on evolution of spiral disks
\citep{roskar08a,roskar08b,schoenrich08,sanchezblazquez09,martinezserrano09,minchev10a,minchev10b}.
The effects of the secular evolution of galaxies are expected to be
most prominent in the outermost parts of disks \citep{roskar08b}.
Additionally, due to their long dynamical timescales, the outer
regions of spirals have largely retained fossil record from the epoch
of galaxy assembly in the form of spatial distributions, kinematics,
ages and metallicities of their stars.  These factors make outskirts
of spirals particularly useful for testing models of galaxy formation
and evolution.

Predictions that models of secular evolution of spirals make for
stellar ages and metallicities in disk outskirts
\citep{roskar08a,roskar08b,sanchezblazquez09} are still to be
confirmed in a sufficiently large sample of galaxies.  Age behavior is
very difficult to observe in galaxies too distant for the full star
formation history to be modeled. \citet{barker07} and
\citet{williams09} demonstrate nicely how an age gradient can be
derived over the whole disk of M33; however, this kind of work is only
possible for the few closest spirals.  On the other hand, abundance
gradients in galaxies can be studied relatively easily out to a few
Mpc using broad-band photometry.  It has been well-established that in
a disk which is growing inside-out, metallicity decreases from the
center of the galaxy outward
\citep{goetzkoeppen92,matteuccifrancois89}.  However, an increasing
number of galaxies show signs of an abundance gradient flattening in
their outermost regions
\citep{andrievsky04,yong06,carraro07,pedicelli09,bresolin09,vlajic09}.
While models mentioned above offer possible explanations for the
flattening, its exact cause is not yet clear and abundance data on
more galaxies of different sizes and outer disk structure will help
explain the slope of abundance gradient in the outermost regions of
spirals.

As discussed in our earlier papers, the Sculptor Group is an ideal
laboratory for studies of pure spirals: (i) its high galactic latitude
minimizes foreground extinction and contamination from Milky Way
stars, (ii) most of its galaxies are isolated disk systems, and (iii)
the distance of the Sculptor Group ($2-4$ Mpc) allows for the resolved
stellar population studies from the ground.  At the far side of the
Sculptor Group ($\sim3.6$ Mpc), NGC~7793 is a late-type spiral with an
absolute magnitude $M = -18.46$, slightly brighter than NGC~300
($-18.12$; absolute magnitudes were derived using corrected total
apparent magnitude from \citet{carignan85} and the distance derived in
\S\ref{sec:trgb}).  \citet{carignan85} studied the light profile of
NGC~7793 out to $6'$ using surface photometry and derived a scale
length of $1.10'$ (1.16 kpc) in $B$.  A neutral hydrogen study of
\citet{carignanpuche90} reveals an \HI disk extending out to $11.2'$
(11.8 kpc), with a surface density at the last data point of
$\sim0.01$ M$_{\odot}$ pc$^{-2}$.

The plan of the paper is as follows.  In \S\ref{sec:obs_dr_phot} we
describe the observations, data reduction, photometry and completeness
analysis.  Next, we present results for the distance of NGC~7793 as
estimated from the tip of the red giant branch (TRGB,
\S\ref{sec:trgb}), color-magnitude diagram (CMD, \S\ref{sec:cmd}),
star counts profiles (\S\ref{sec:profile}), surface brightness profile
(\S\ref{sec:surfacebrightness}) and metallicity gradient
(\S\ref{sec:mdf}).  Discussion and conclusions follow in
\S\ref{sec:discussion}.

\section{OBSERVATIONS, DATA REDUCTION AND PHOTOMETRY}
\label{sec:obs_dr_phot}

\subsection{Observations and Data Reduction}
\label{sec:observations}

\begin{figure}[t]
\epsscale{1.1}
\plotone{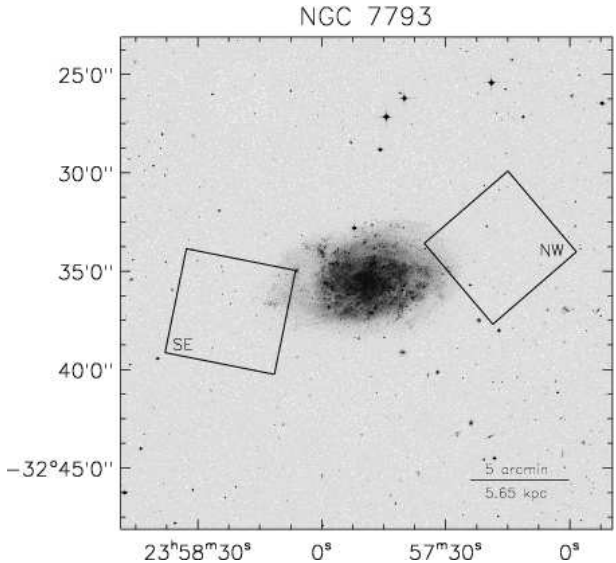}
\caption{DSS wide-field image of NGC~7793 with GMOS fields overlaid.
\label{fields}}
\end{figure}

The data were obtained using the Gemini Multi Object Spectrograph
(GMOS) on the Gemini South telescope over three nights in 2005 August
as a part of the program GS-2005B-Q-4.  Deep $g'$ and $i'$ images of
two major axis fields on each side of the galaxy were taken; the
locations of the fields (SE and NW) are shown in Figure~\ref{fields}.
GMOS field of view is $5'.5$ on a side.  (At the distance of NGC~7793
(3.61 Mpc, \S\ref{sec:trgb}), $1'$ corresponds to $1.05$ kpc.)  The
average FWHM of the data is $0''.6$ (SE) and $0''.8$ (NW) in $g'$, and
$0''.5$ (SE) and $0''.7$ (NW) in $i'$ band.

To reduce the data we employed the standard IRAF/Gemini routines which
included (i) bias subtraction and flat fielding (\texttt{gireduce}),
(ii) (for $i'$ data only) creating the master fringe frame from the
individual reduced frames (\texttt{gifringe}) and subtracting it from
the individual images (\texttt{girmfringe}), (iii) mosaicking of
individual GMOS CCDs into a single reference frame (\texttt{gmosaic}),
and (iv) combining the dithered exposures into a final image
(\texttt{imcoadd}).

We obtained $13\times600$ $s$ exposures in $g'$ and $22\times600$ $s$
exposures in $i'$ band per field, bringing total on-source exposure
time to $11.7$ hours.  The data for the SE field were taken during the
nights of 2005 Aug 09UT (hereafter: first night) and 2005 Aug 10UT
(hereafter: second night).  Only 5 (out of 13) $g'$-band and 21 (out
of 22) $i'$-band images were observed on the first night and the
observing log indicated that a thin cloud might have affected
$i'$-band observations.  In addition, only one standard star field
observation was recorded.  Remaining science images of SE field, as
well as three standard stars fields were observed the following night.
The bulk of the data of the NW field was collected the night of 2005
Aug 11UT (12 $g'$-band and 21 $i'$-band images; hereafter: third
night).  However, no photometric standard stars were observed that
night.  The remaining science frames (one in each of the bands) were
observed the previous night under photometric conditions.

Initial analysis of the photometry revealed a suspicious discrepancy
in $i'$-band magnitude distribution between the two fields.  In
correspondence with the Gemini staff it was confirmed that this was
most likely due to the non-photometric conditions on the night of 2005
Aug 09UT, when the majority of the science frames of the SE field were
taken.  Accounting for this, and the fact that the photometric
standard stars observation were only taken on the second night of the
run, we decided to proceed in the following manner:

- SE field, $i'$-band: combine 21 images taken on the first night and
  compare photometry of this deep image with the photometry extracted
  from the single image observed on the second night.

- SE field, $g'$-band: combine separately 5 images observed on the
  first, and 8 images taken on the second night and compare the
  photometry between the two images.

- NW field, $i'$-band: compare the photometry extracted from the deep
  image created by combining 21 images taken on the third night of the
  run with the single image taken on the second night.

- NW field, $g'$-band: similarly to the $i'$-band case, compare a
  combined deep image created from the 12 exposures taken on the third
  night, with the single image taken on the second night of the run.

In the following sections, we will refer to the (combinations of)
images taken on the second night of the run as calibration images and
to the ones observed on the first and third night as final images.

\subsection{Photometry and Artificial Stars Tests}
\label{sec:photometry}

To extract stellar photometry we used the standalone version of
DAOPHOT and ALLSTAR packages \citep{stetson87}.  Following the initial
runs of \textsc{find} and \textsc{photometry} routines, which were
used to catalog objects in the image and measure their aperture
photometry, we proceeded to determine the point spread function (PSF)
for each image. Depending on the filter and the field, $80-220$
moderately bright isolated stars in each field were selected as PSF
stars and used to iteratively compute the PSF.  The PSF stars were
``hand-picked'' and their radial, contour and mesh profiles were
visually examined within the IRAF/DAOPHOT package.  The calculated
point spread function was used to subtract the PSF stars from the
original image; the positions of the subtracted PSF stars were
inspected again and the PSF stars which did not subtract cleanly were
excluded from the PSF calculation.  In addition, stars with
subtraction errors which differed more than $3\sigma$ from the mean
value were also excluded.  The next iteration of the point spread
function was calculated using images in which, within the fitting
radius of each PSF star, all but PSF stars have been subtracted.  This
was followed with yet another visual inspection within IRAF/DAOPHOT as
described above.  The whole procedure was repeated once more to derive
the final PSF.  Finally, ALLSTAR was used to fit the calculated PSF to
all stars in the object catalogs and determine their photometry.

Comparison of the final and calibration images revealed the following.
In the $i'$-band, we found the difference in photometry extracted from
the calibrated and final images of $0.396\pm0.012$ for the SE field
and $0.073\pm0.006$ for the NW field.  In the $g'$-band we found,
within uncertainties ($2\sigma$), no difference in magnitudes between
calibration and final images.  We therefore decided, in the subsequent
analysis, to use the master frames created by combining the full
sample of 13 frames in $g'$, and a reduced sample of 21 images in the
$i'$-band.  Two $i'$-band images, one of each field, taken on the
second night were used to calibrate the photometry of the final
images which were very likely taken under non-photometric conditions.

We performed photometric calibration using the standard stars
observations taken during the second night of the observing run.  The
derived zero points were in excellent agreement with those stated on
the Gemini website (28.31 (our value) vs. 28.33 (GMOS value) for $g'$
and 27.92 (our value) vs. 27.93 (GMOS value) for $i'$ band).  In
addition, we applied additional offset (0.396 for the SE field, 0.073
for the NW field) to the photometry derived from our final $i'$
images, as explained above.

The DAOGROW software suite \citep{stetson90} was used to calculate
aperture corrections.  We subtracted from each image all but PSF stars
and determined their aperture photometry in a series of apertures.
DAOGROW uses the information on aperture photometry to derive the
``total'' magnitude of a star.  The weighted mean difference between
the PSF-based magnitude (from ALLSTAR) and the ``total'' magnitude
(from DAOGROW) of the PSF stars was adopted as 'aperture correction'
and applied to PSF-based magnitudes of all stars in the frame. The
uncertainties in the aperture correction were in the range
0.006-0.008.

Completeness, crowding and photometric uncertainties of the data were
assessed using artificial star tests, described in more detail in
\citet{vlajic09}.  We added $900$ artificial stars to each frame in a
$30\times30$ grid with the cell size of $69$ pixels ($10''$); for each
combination of field and filter $100$ artificial star test runs were
performed with the grid origin randomly offset between the runs.
Photometry (i.e. magnitude range and color distribution) of artificial
stars was determined by randomly sampling the original color magnitude
diagrams.  Artificial stars were added to the frames using the
DAOPHOT/ADDSTAR routine and the resulting images were analyzed with
the identical data reduction pipeline as the original frames.  We
consider an artificial star recovered if it is detected in both
filters with the difference between input and recovered magnitude
$<0.5$ mag.  Completeness as function of magnitude is calculated as a
ratio between the number of recovered and input stars in a given
magnitude bin.  Our photometry is $50\%$ complete down to
($g'$,$i'$)=($27.35$--$27.47$,$26.75$--$27.10$).  We also calculate radial
completeness as a ratio of the number of recovered and input stars in
a given radial bin; the information is later used to correct radial
star counts and effective surface brightness profiles for crowding in
the innermost regions.  Furthermore, we used information on the
\textsc{chi} and \textsc{sharp} parameters of recovered artificial
star tests to discard spurious objects from the ALLSTAR
catalog. \textsc{chi} and \textsc{sharp} are often employed as
indicators of the non-stellar nature of an object; outliers in the
plots of \textsc{chi} and \textsc{sharp} as a function of magnitude
can be relatively safely assumed to be semi-resolved galaxies, blends,
cosmic rays or image defects.  We removed from further analysis all
stars lying outside the envelope demarcating $3$ standard deviations
from the artificial stars' mean \textsc{chi} and \textsc{sharp}
values.  These cuts, together with a cut on error in color
($\approx0.15$) remove $\sim50\%$ of objects from the original matched
catalogs.

\section{RESULTS}
\label{sec:results}

\subsection{Tip of the Red Giant Branch Distance}
\label{sec:trgb}

Stellar evolution models
\citep[e.g.][]{salariscassisi97,ibenrenzini83} show that below a
metallicity of [Fe/H]$\approx-0.7$ the brightness of the tip of the
red giant branch (TRGB) is expected to be roughly constant, regardless
of stellar age and metallicity.  Observationally, \citet{dacosta90}
and \citet{bellazzini01} confirmed that for a sample of Galactic
globular clusters spanning a large range of abundances
($-2.1<[\mathrm{Fe/H}]<-0.7$) and ages ($2-15$ Gyr), the absolute
$I$-band magnitude of the TRGB is stable and insensitive to age and
metallicity, which enables its application as an extragalactic
distance estimator.  In general, there is been a good agreement
between the distances obtained using the TRGB and classical methods
such as Cepheids \citep{sakai96,ferrarese00}.

\begin{figure}[!ht]
\epsscale{1.1}
\plotone{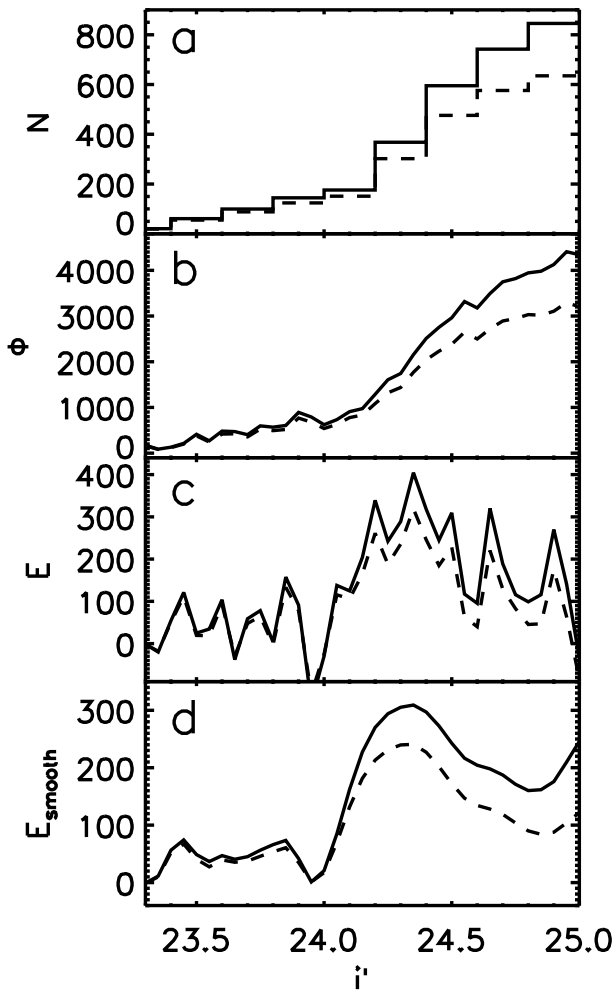}
\caption{Determination of the tip of the red giant branch.  $i'$-band
luminosity function (a) and smoothed luminosity function (b) are shown
in top panels before and after completeness correction (dashed and
solid lines respectively).  Outputs of the edge-detection filter are
shown in the bottom two panels -- calculated using the photometric
error as determined from artificial star tests (c), and calculated
using a photometric error four times larger than calculated from
artificial star tests (d); line types are the same as in the top
panels.\label{trgb}}
\end{figure}

Following \citet{lee93} and \citet{sakai96}, we employ our deep
$i'$-band photometry of two fields in the outskirts of NGC~7793 to
determine the galaxy's distance.  Figure~\ref{trgb} shows the result
of this exercise.  In order to increase the sample of stars used in
the detection, we combined photometry of the two fields.  This results
in a value consistent with the one we derive from each field
separately, while having a narrower and better defined peak.

The top panel of the Figure shows the $i'$ band luminosity function,
with (solid line) and without (dashed line) the completeness
correction.  In correcting for incompleteness, we only consider stars
that are above $50\%$ completeness limit in both $g'$ and $i'$.  In
the middle panel, the smoothed luminosity function is shown.  In
calculating this function, discrete stellar magnitudes are replaced by
Gaussians with the width equal to the star's photometric error,
according to the expression \citep{sakai96}:

\[ \Phi(m)=\sum_{i=1}^{N}\frac{1}{\sqrt{2\pi}\sigma_{i}} 
\mathrm{exp}\bigg[-\frac{(m_i-m)^2}{2\sigma_i^2}\bigg] \]

A smooth adaptive edge-detection filter of the form:

\[ E(m) = \Phi(m+\overline\sigma_m)-\Phi(m-\overline\sigma_m) \]

\noindent is then applied.  Here, $\overline\sigma_m$ is the mean
photometric error of all stars in the magnitude range
$[m-0.05,m+0.05]$, as determined from artificial star tests.  The
maximum of the output of this edge-detection filter (shown in
Figure~\ref{trgb}c) marks the TRGB.  Relatively small photometric
errors cause the output of the edge-detection filter to be noisy.  In
order to obtain a smoother output which would yield a more reliable
distance estimate, we calculate the same edge-detection function using
photometric errors three times those estimated from the artificial
star tests.  The result is shown in the bottom panel of
Figure~\ref{trgb}.  Apparent $i'$ magnitude of TRGB is estimated to be
$i'_{TRGB}=24.35\pm0.30$. Uncertainties are determined as
half-width-half-maximum of the peak in Figure~\ref{trgb}d.  With the
absolute $i'$-band magnitude $M_{i',TRGB}=-3.44\pm0.10$
\citep{bellazzini08}, we calculate the NGC~7793 distance modulus to be
$m-M=27.79\pm0.32$, corresponding to the distance of $3.61\pm0.53$
Mpc.  This is in good agreement with the values obtained by
\citet[][$3.91\pm0.41$ Mpc]{karachentsev03} and the GHOSTS survey
\citep[$3.73$ Mpc][]{radburnsmith11a}.

\subsection{Color-Magnitude Diagram}
\label{sec:cmd}

\begin{figure}[t]
\epsscale{1.1}
\plotone{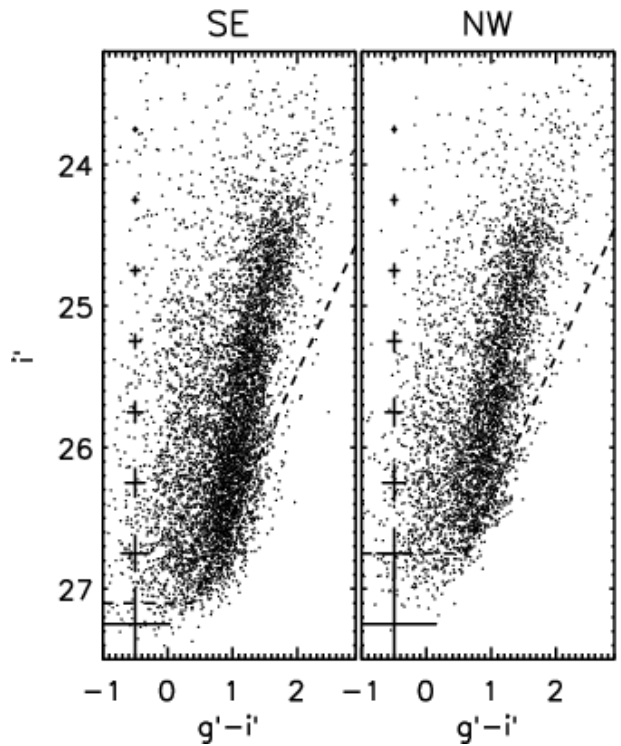}
\caption{Color-magnitude diagram of two NGC~7793 fields.  Dashed lines
represent the $50\%$ completeness limit.  Photometric uncertainties at
a given magnitude and color of $g'-i'=1$ are shown on the left of the
panels.\label{cmd}}
\end{figure}

Color-magnitude diagrams (CMDs) of two NGC~7793 are shown in
Figure~\ref{cmd}.  Also marked in the Figure are $50\%$ completeness
limit and photometric errors as determined from artificial star tests
at a given $i'$ magnitude and color of $g'-i'=1$.  The ($g'-i'$,$i'$)
CMDs reveal a prominent red giant branch (RGB).  RGB stars make up
more than $60\%$ of detections in each field.  Region just blueward of
the RGB is most likely populated by unresolved faint background
galaxies.  Also potentially visible in the CMD is a main sequence (MS)
and an asymptotic giant branch (AGB) population.  Selection boxes used
for deriving properties of individual stellar populations are shown in
Figure~\ref{cmdboxes}.  AGB and MS populations make up $\sim2\%$ and
$\sim10\%$ of all objects in the field.

\subsection{Star Counts Profiles}
\label{sec:profile}

We derive radial profiles of NGC~7793 in order to examine the extent
and structure of its outer disk.  Star counts profiles are computed by
counting up the stars in elliptical annuli with inclination and
position angle corresponding to that of NGC~7793.  These raw counts
are then normalized by the area sampled by each annulus.

The most challenging aspect of studying outer disk light profiles is
the need to determine a background level that is then subtracted from
the raw profile in order to calculate the true star counts or surface
brightness profile of a galaxy.  In the case of surface photometry
this is manifested through difficulties in deriving accurate estimates
of sky brightness.  When resolved stellar photometry is used instead
to study outskirts of spirals, the problem translates into how to
reliably evaluate the number of unresolved faint background galaxies
that are mistakenly included into stellar catalogs.  Unlike in the
case of surface photometry, in star counts studies we are not limited
by the fundamental limits of observations but by the accuracy with
which we can determine the galaxy number counts.  However, the
challenge is particularly difficult when information on number counts
as a function of color (and not only magnitude) is required.  We
therefore pay careful attention to estimating the number counts of
faint background galaxies.

\begin{figure}[t]
\epsscale{1.1}
\plotone{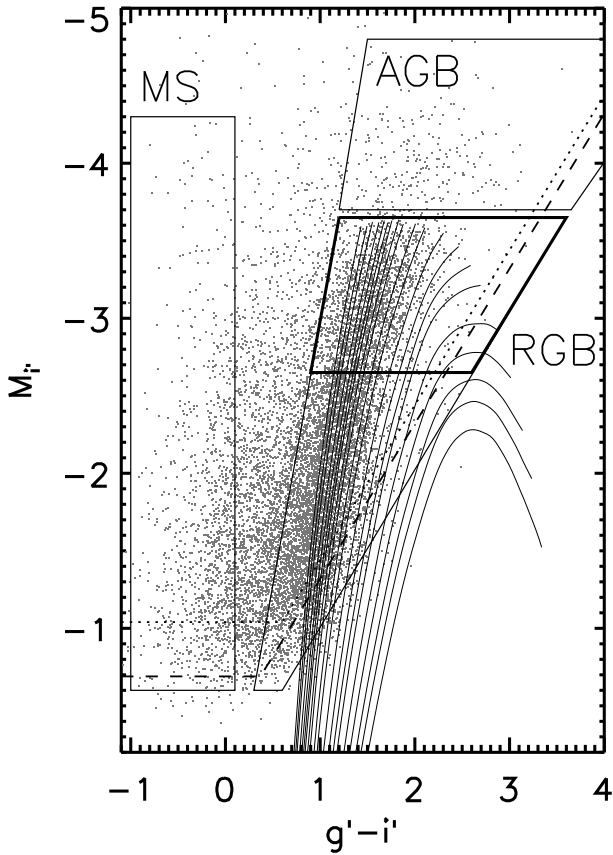}
\caption{Color-magnitude diagram of both NGC~7793 fields with
selection boxes for RGB, AGB, and MS populations marked.  Dashed lines
represent the $50\%$ completeness limit for SE field (dashed line) and
NW field (dotted line). \label{cmdboxes}}
\end{figure}

\begin{figure*}[t]
\epsscale{0.8}
\plotone{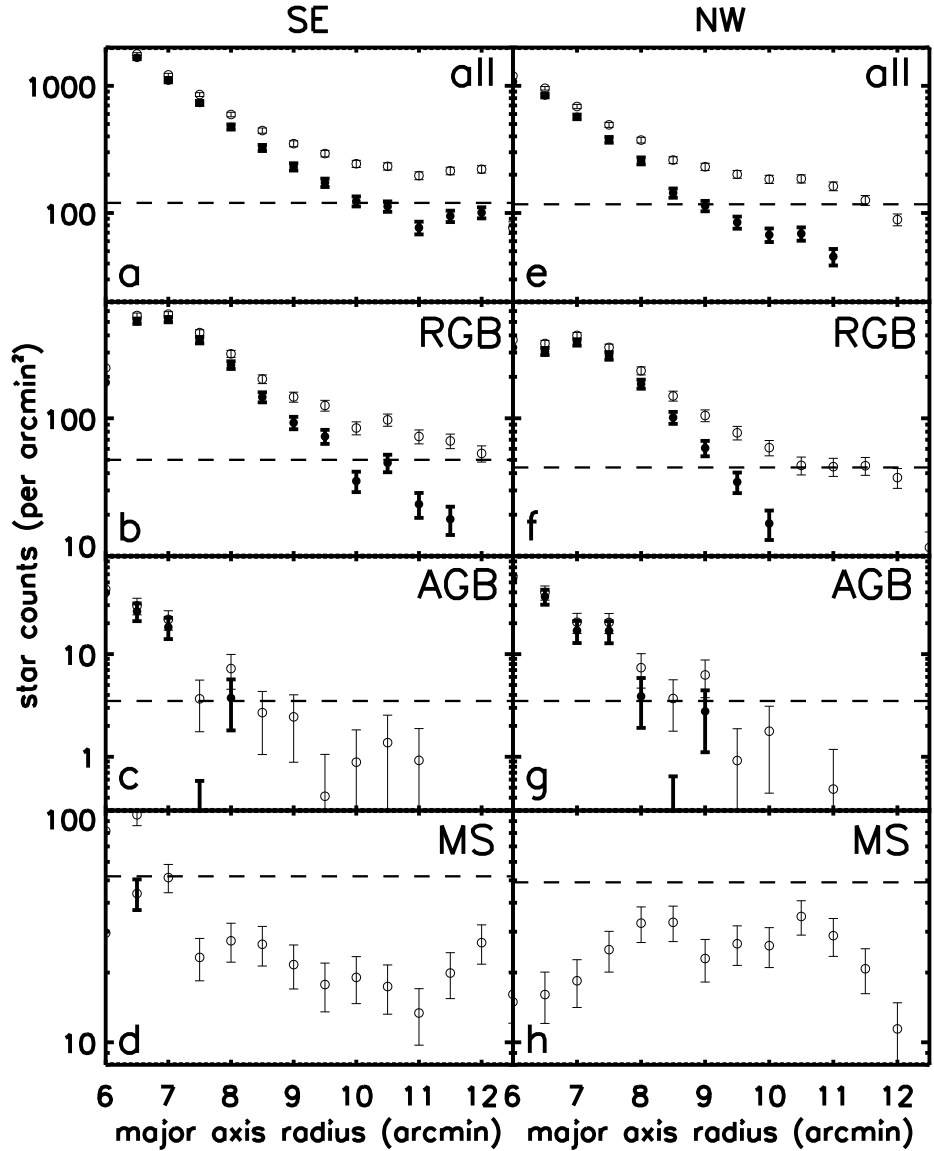}
\caption{Starcounts profiles of the SE field (left panels) and NW
field (right panels).  From top to bottom, the profiles shown are of
all stars (a \& e), RGB (b \& f), AGB (c \& g) and MS stars (d \&
h). In each panel the dashed line displays the estimated galaxy counts
level, determined using methods described in the text.  Open and full
symbols show the star counts before and after the background galaxy
contribution has been subtracted.
\label{profile}}
\end{figure*}

In Figure~\ref{profile}a,e we show star count profiles constructed
using all stars in the field, irrespective of their colors.  We use
the {\sl GalaxyCount} online tool \citep{ellisbh07} to determine
galaxy number counts in the magnitude range probed by our data; dashed
horizontal line in top panels shows the estimated background galaxy
level of $120(\pm12)$ arcmin$^{-2}$ (SE) and $117(\pm12)$
arcmin$^{-2}$ (NW field).  The open and full symbols in the figure
represent the star counts before and after the number counts of galaxy
contaminants have been subtracted.  Figure~\ref{profile}e shows that
the star counts profile of the NGC~7793 NW field falls off to
practically zero within the extent of our data, at the radius of
approximately $11'$ or $11.5$ kpc.  It is unclear whether the upturn
reflected in the two outermost datapoints for the SE field
(Figure~\ref{profile}a) is real, or (possibly more likely) a result of
a slight underestimation of background counts.

We also calculate the star counts profiles of distinct stellar
populations in the outer disk of NGC~7793
(Figure~\ref{profile}b-d,f-h).  Selection boxes used to separate stars
into different populations are marked on the color-magnitude diagram
in Figure~\ref{cmdboxes}.  Since the {\sl GalaxyCount} tool does not
provide direct information on the color of background galaxies, we
require an alternative method for determining the galaxy number
counts.  Following \citet{vlajic09}, we employ the data from the
William Herschel Deep Field \citep[WHDF,][]{metcalfe01} to estimate
the contamination from the faint background galaxy population.  We
determine the number counts within the asymptotic giant branch (AGB)
box directly from the WHDF data, while for the stellar populations
reaching fainter magnitudes than probed with WHDF (i.e. main sequence
(MS) and red giant branch (RGB) stars) we use the method described in
\citet{vlajic09} to calculate the galaxy number counts.  We calculate
the $i'$-band number counts of all galaxies in WHDF and fit linearly
the (log of) differential number counts in $0.5$ mag bins.  In order
to determine the galaxy counts below the limit of the WHDF survey we
assume that the counts in the bins $2-3$ magnitudes below the survey
limit follow the same linear trend (in the log space) as the counts in
the brighter bins used in the fit.  We finally correct the derived
galaxy number counts using completeness curves of our data.  The
resulting background galaxy counts are $3.9\pm2.1$, $51\pm7$
($44\pm7$) and $52\pm8$ ($49\pm7$) arcmin$^{-2}$, for the AGB, RGB and
MS selection regions, respectively, for the SE (NW) field.  (Quoted
errors are variance, as estimated by {\sl GalaxyCount}.)  While
contamination-subtracted profiles of RGB stars largely confirm the
finding from Figures~\ref{profile}a,e (with the distinction that the
RGB profile for the NW field falls off more steeply and RGB stars are
only detected out to $10'$) we detect no main sequence stars and all
objects within our MS selection box can be attributed to the
contaminating background galaxy population (the galaxy number counts
for the MS selection box are $\sim2-3$ times higher than the derived
star counts for this color-magnitude region).  We detect AGB stars out
to $8-9'$ ($8.5-9.5$ kpc), after which their number counts fall bellow
the estimated background galaxy level.  As we show in
\citet{vlajic09}, at the high galactic latitudes of the Sculptor
Group, contamination from the Milky Way stars is negligible
\citep{robin03,sharma11}.

\citet{radburnsmith11b} find a break in the radial profile of young
and intermediate age stars in the outer disk of NGC~7793 (their
HST/ACS fields overlap significantly with our SE field), with the
scale length of a stellar population being shorter for younger stars.
This is largely consistent with the star counts profiles we derive.
Due to the higher level of contamination in our ground based data we
see no MS stars, in agreement with the short scale length for this
population found by \citet{radburnsmith11b}; similarly, we find AGB
stars to be more extended than the MS population, with the RGB stars
having the largest scale length.

Comparing the RGB profiles of NGC~7793 and NGC~300
\citep[Figure~9]{vlajic09} we find that the counts in the outermost
bins shown in Figure~\ref{profile}b are $\sim2$ times lower than
corresponding counts in the most distant bins in the outer disk of
NGC~300.  This is yet another piece of evidence supporting our earlier
finding of an extended exponential disk in NGC~300
\citep{blandhawthorn05,vlajic09}.  While our CMD of NGC~300 reaches 4
mag below the tip of the RGB, compared to only 2.5 mag in NGC~7793,
background galaxy number counts increase rapidly with magnitude and
the counts in the faintest magnitude bins dominate the total galaxy
counts.  Our CMDs of NGC~300 and NGC~7793 reach same apparent depth
($\sim26.5-27$ mag) and hence experience roughly the same
contamination by faint background galaxies.  The difference in star
counts in the outermost bins therefore does not reflect the difference
in galaxy number counts but in star counts, and is an additional
independent confirmation of the extended exponential disk in NGC~300
out to at least $10$ disk scale lengths.

\subsection{Surface Brightness Profile}
\label{sec:surfacebrightness}

The power of resolved stellar photometry over surface photometry is
most easily recognized if star counts are transformed into
measurements of effective surface brightness and compared with
existing surface brightness data.  It has been shown in a number of
works recently \citep{blandhawthorn05,irwin05,dejong07,radburnsmith11a}
that this approach allows one to reach surface brightnesses $3-4$ mag
arcsec$^{-2}$ below the limit of surface photometry.

We divide the data in $0.5'$ wide annuli and calculate surface
brightness in each annulus as:

\[ \mu = -2.5\log\Big(\frac{F''}{N_{pix}}f_{g'}f_{i'}
\sum_{i}10^{-0.4m_{i}}\Big) \]

Here, $N_{pix}$ is a number of pixels in an annulus, $f_{g'}$ and
$f_{i'}$ are radial completeness factors and $m_{i}$ are magnitudes of
stars within a given annulus.  Radial completeness of our data is
lowest in the innermost annulus ($45\%$) due to crowding, and
increases to an average of $87\%$ in the outermost disk.  To convert
surface brightness to units of mag arcsec$^{-2}$ we multiply the
effective flux under the logarithm with the inverse of the square of
the GMOS pixel size ($1$ pix$=0.146''$) which is equal to
$F''=1/0.146^2=47$.

In order to take into account the contribution from faint background
galaxies (see \S\ref{sec:profile}) we estimate contamination in each
annulus as 

\[ N_{cont}=\frac{N_{gal}N_{pix}}{F'} \] 

\noindent where $F'=168887$ is a scale factor equivalent to $F''$ that
converts $N_{pix}$ to arcmin$^{2}$ and $N_{gal}=120(\pm12)$
arcmin$^{-2}$ for SE and $117(\pm12)$ for NW field
(\S\ref{sec:profile}).  We next perform 100 realizations of background
subtraction; in each of these we remove a random set of $N_{cont}$
stars from each annulus and calculate surface brightness using the
remaining stars as described above.  The final surface brightness
profile is the mean of a hundred realizations.  We find the standard
deviation of the whole set of realizations for a given annulus not to
be significant ($<0.03$ mag arcsec$^{-2}$ in $g'$ and $<0.05$ mag
arcsec$^{-2}$ in $i'$) confirming that the derivation of surface
brightness profile is robust against the choice of objects we exclude
as contamination.

\begin{figure*}[!th]
\epsscale{1.1}
\plottwo{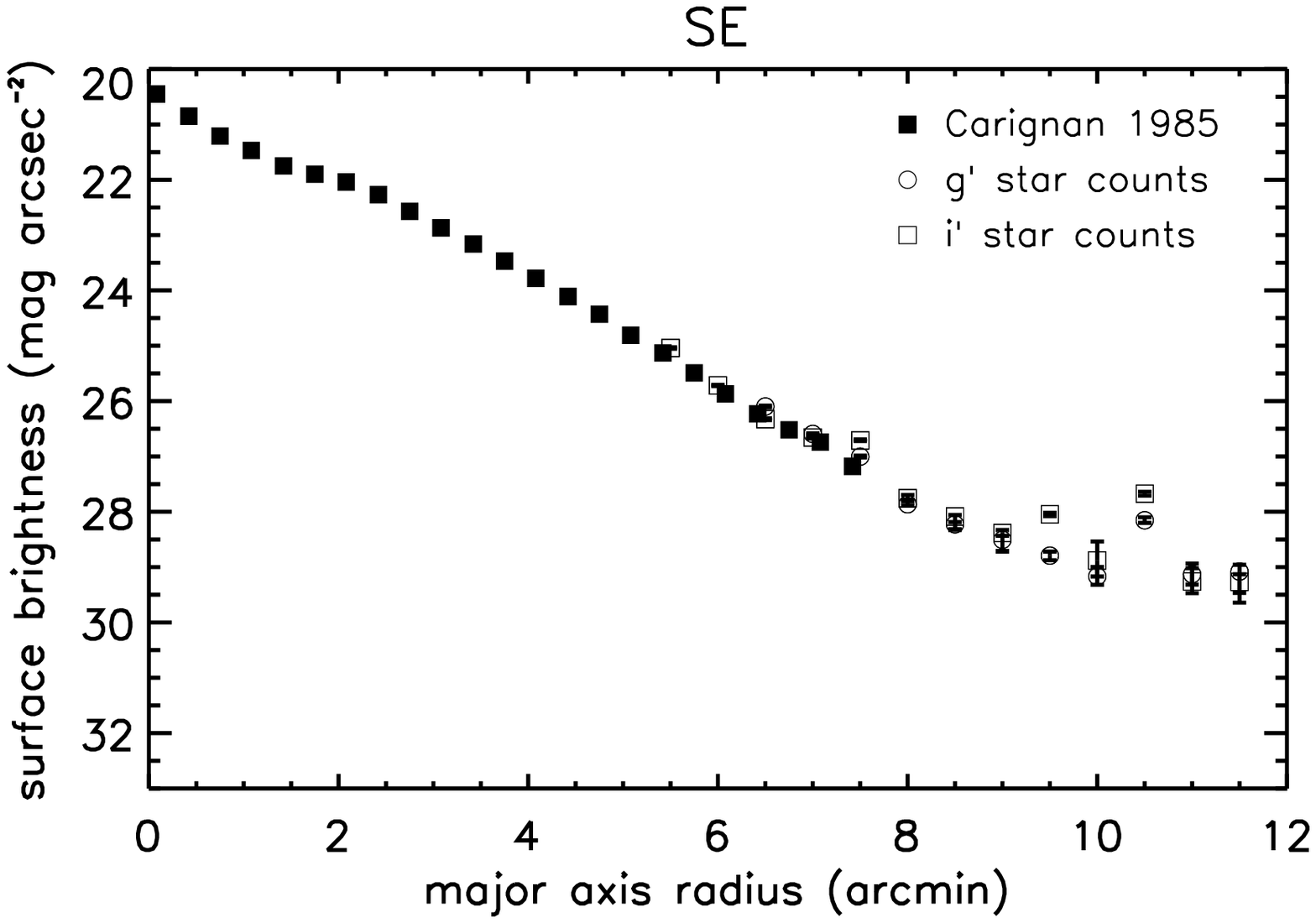}{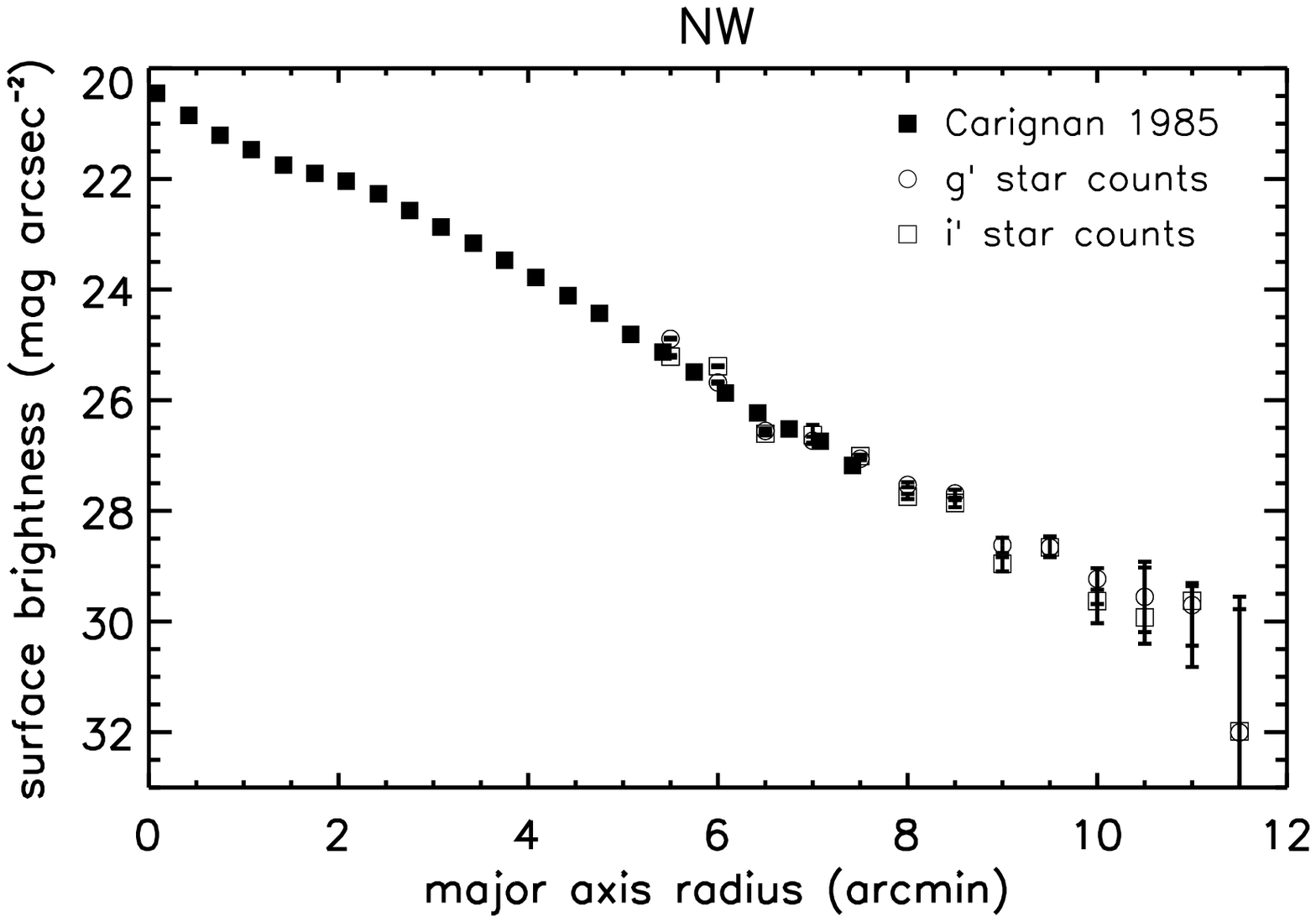}
\caption{Surface brightness profile of NGC~7793.  Inner disk data
(full squares) are surface brightness measurements from
\citet{carignan85}. Empty circles (squares) are effective surface
brightness measurements in $g'$ ($i'$) band, derived from star counts
and corrected for background galaxy contamination.  Error bars reflect
uncertainties in galaxy number counts.  Due to different relative
depths and completeness, data had to be shifted by different amounts
to match the \citet{carignan85} measurements; see text for details.
Resolved stellar counts reach more than $4$ mag arcsec$^{-2}$ deeper
than surface photometry and significantly increase the known extent of
NGC~7793.
\label{surfacebrightness}}
\end{figure*}

The derived $g'$ and $i'$ effective surface brightness profiles are
shown in Figure~\ref{surfacebrightness}.  The inner disk data points
in the figure are from the \citet{carignan85} study.  Since, to
minimize spurious detections, we used the matched stellar catalog
rather than individual $g'$ and $i'$ catalogs when deriving surface
brightness, it is not surprising to find that $g'$ and $i'$ profiles
are practically identical.  The uncertainty in the galaxy number
counts is reflected in the error bars in
Figure~\ref{surfacebrightness}.  Low/high error bar limits represent
surface brightness profiles calculated using background galaxy number
counts that are $1\sigma$ lower/higher than number counts used to
derive the original profile.  In order to match the inner disk result,
effective surface brightness data points of the NW field have been
shifted downward by $1.3$ and $0.3$ mag in $g'$ and $i'$,
respectively.  For the SE field, corresponding values are $0.9$ and
$0.0$ in $g'$ and $i'$.

\begin{deluxetable}{cccc}
\tablecaption{NGC 7793 disk scale length. \label{scalelength}}
\tablewidth{0pt} 
\tablehead{ 
\colhead{Field} & \colhead{Filter} & \colhead{scale length [']} & \colhead{scale length [kpc]}}
\startdata 
Carignan & B$_J$ & 1.11$\pm$0.01 & 1.17$\pm$0.02 \\
SE & g' & 1.12$\pm$0.05 & 1.18$\pm$0.05 \\
SE & i' & 1.24$\pm$0.09 & 1.30$\pm$0.10 \\
NW & g' & 1.12$\pm$0.07 & 1.18$\pm$0.07 \\
NW & i' & 1.05$\pm$0.06 & 1.10$\pm$0.07
\enddata 
\end{deluxetable}

Fitting an exponential to a light profile in
Figure~\ref{surfacebrightness} we calculate the scale length of the
NGC~7793 disk.  \citet{carignan85} data points yield for the disk
scale length a value of $1.11'\pm0.02$ ($1.17\pm0.02$ kpc).  The scale
lengths derived using our data, excluding data points at $10.5$ and
$11'$, are presented in Table~\ref{scalelength} and are largely in
agreement with the Carignan's value.

In summary, our effective surface brightness derived from star counts
reaches $\sim3$ mag arcsec$^{-2}$ deeper than the surface brightness
profile of \citet{carignan85}.  We trace the exponential light profile
out to $\sim9$ disk scale lengths, greatly increasing the known extent
of the disk from the Carignan study.  There is no detectable break in
the surface brightness profile and the disk remains exponential out to
$11'$ ($11.5$ kpc).

\subsection{Metallicity Distribution Function and Metallicity Gradient in the Outer Disk}
\label{sec:mdf}

Using the position of stars on the RGB as a proxy for their
metallicities, we derive the metallicity distribution function (MDF)
and the metallicity gradient of the stars in the outer disk of
NGC~7793.  Due to higher sensitivity of the color of RGB stars on
metallicity than age \citep[e.g.][]{vandenberg06}, the age-metallicity
degeneracy does not have a significant effect on the shape of the
derived MDF and metallicity gradient, and allows for a relatively
accurate abundance determination from broad band photometry.

In calculating the metallicity gradient, we only use stars from the
top $\sim1$ magnitudes of the RGB selection box in
Figure~\ref{cmdboxes} (i.e. with $M_i'<-2.5$).  At fainter magnitudes
stars at the red/metal-rich side of the RGB fall below the
completeness limit, and the resulting MDF is artificially skewed
towards the metal-poor end.  We adopt four different sets of stellar
evolutionary tracks from \citet{vandenberg06} (with ages of 8 and 12
Gyr and [$\alpha$/Fe] of $0.0$ and $0.3$) and interpolate between them
to derive metallicities on a star by star basis.  The model grid
consist of 16 finely spaced red giant tracks covering the range of
metallicities from [Fe/H]$= -2.31$ to $0.00$ in the steps of
approximately $0.1$ dex.

\begin{figure*}[t]
\epsscale{1.1}
\plottwo{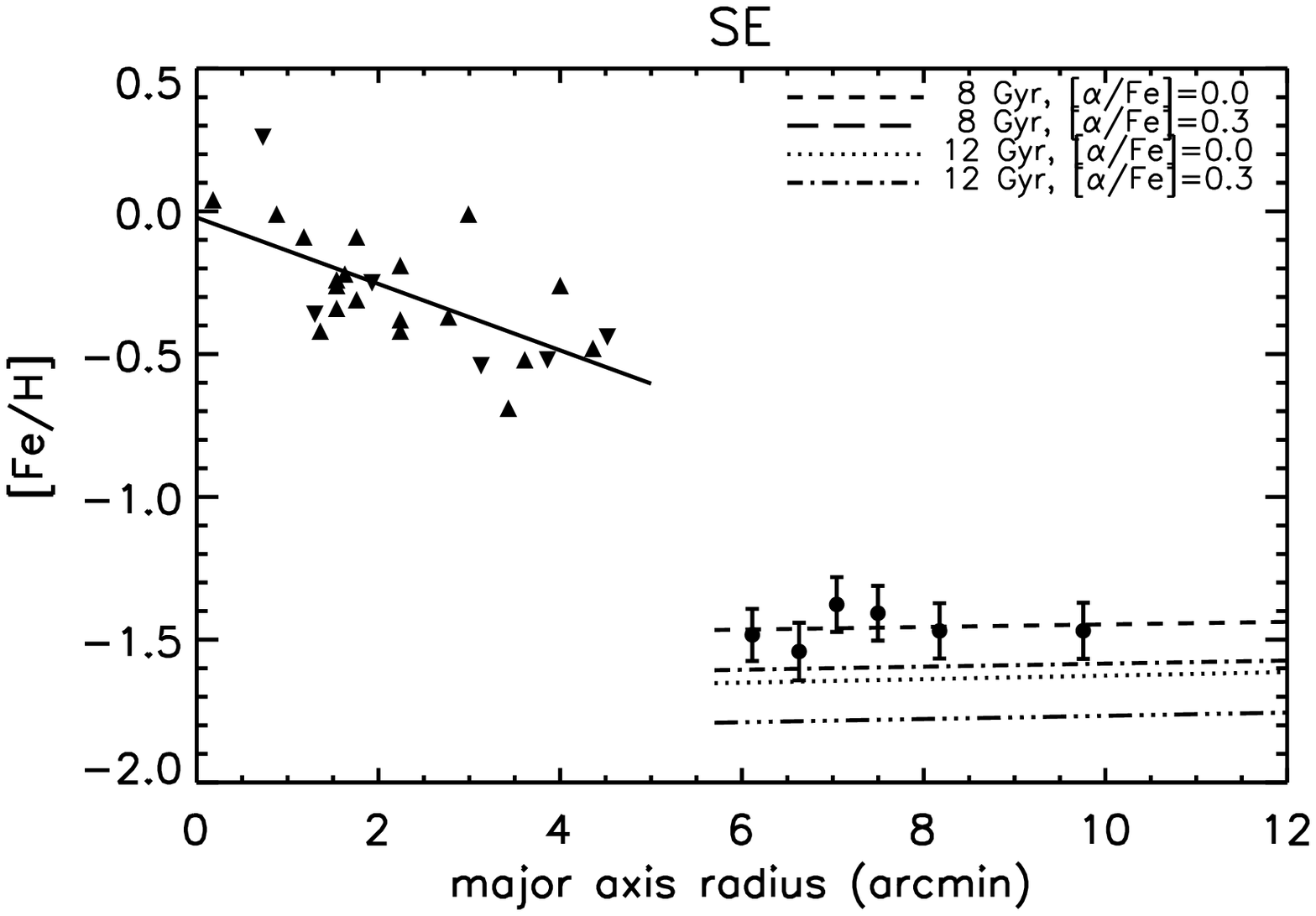}{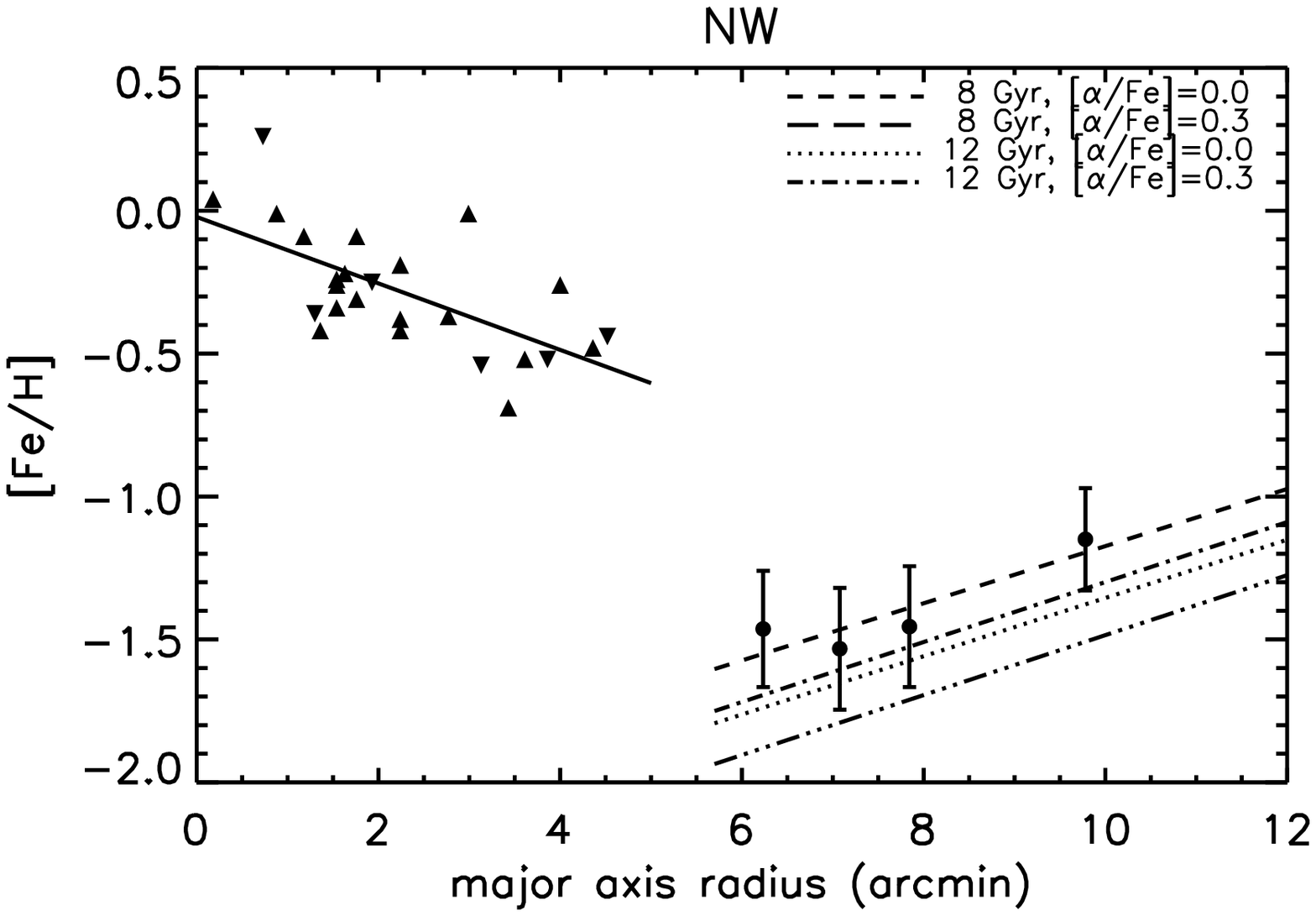}
\caption{Metallicity gradient in NGC~7793.  Filled triangle symbols
are inner disk [O/H] abundances from the works of
\citet{edmundspagel84} and \citet{webstersmith83}; the solid line
shows a linear fit to these abundances.  Outer disk metallicity
profiles are linear fits to radially binned metallicities presented in
this work.  Different line styles denote abundance profiles derived
using different isochrone sets; ages and [$\alpha$/Fe] ratios of these
isochrones are shown in the top right.  For clarity, we only show
binned metallicities that were used to derive the abundance profile
for one isochrone set (filled circles).  Error bars represent mean
value of abundance uncertainties in each bin.  \label{gradient}}
\end{figure*}

We divide the data into bins of at least 200 stars and for each
isochrone set determine median metallicities in these bins.  The
metallicity gradient is calculated as a slope of a linear fit to the
binned metallicities.  Effects of using stellar tracks with different
ages and/or $\alpha$-enhancements are shown in Figure~\ref{gradient}
and Table~\ref{alphafe_age}.  The use of older and more
$\alpha$-enhanced isochrones results in lower overall metallicities,
but has practically no effect on the derived abundance gradient.

Figure~\ref{gradient} reveals two surprising characteristics -- a
strong positive metallicity gradient in the NW field and overall low
metallicities in both fields.  It becomes clear upon closer inspection
of the Figure~\ref{gradient} that the outermost binned datapoint is
primarily responsible for the overall steep gradient in the NW field.
The most distant radial bin covers a radial range of $\sim3'$; a
substantial fraction of detections in these outermost few arcminutes
can be attributed to background contamination (Figure~\ref{profile}f),
making the number count of actual RGB stars very low.  If the most
distant data point is discarded, the abundance gradient becomes flat,
with the slope and overall metallicity consistent with that derived
for the SE field.  However, the value of the slope does not change
with the change of the bin size, suggesting that the positive
abundance gradient in the NW field is real.

On the issue of very low metallicities, we see two possible causes for
such behavior.  Errors in the photometric calibration would manifest
as a displacement of the RGB within the color-magnitude diagram, which
would in turn result in false metallicity estimates.  We reject this
possibility, since due to the initial problems with the photometry
(\S\ref{sec:obs_dr_phot}), photometric calibration was checked
extensively and independently by Gemini personnel, and their analysis
confirmed our zero point values.  Alternatively, it is possible that
the component we are seeing in the outskirts of NGC~7793 is not an
outer disk, but a galactic halo.  However, the effective surface
brightness profile we derive follows the inner disk profile
(Figure~\ref{surfacebrightness}) and it seems very unlikely that the
transition from the disk to the halo occurs already at $6'$, where we
derive stellar metallicities of [Fe/H]$\approx-1.5$.  Moreover,
\citet{radburnsmith11a} derive metallicities in the outer disk of
NGC~7793 and find low mean abundances ranging from $-1.23$ to $-1.64$
in their Fields 01 and 02 (which largely overlap with our SE field).

In order to assess the magnitude of the apparent discontinuity in the
region where the inner and the outer disk metallicities overlap
($\sim6$ arcmin, Figure~\ref{gradient}) it is necessary to take into
account (i) the [$\alpha$/Fe] ratio, which corrects for the fact that
inner and outer disk abundances refer to different elements, and (ii)
the age-metallicity relation, which reconciles abundances in young HII
regions with metallicities of the stars on the red giant branch which
we have assumed to be at least 8 Gyr old.  Despite numerous results
pointing to a flat age-metallicity relationship in the Solar
neighborhood \citep[e.g.][]{edvardsson93}, these works find that the
relation steepens for the oldest stars.  Assuming that the difference
in metallicities of young HII regions and $8-12$ Gyr old RGB stars is
$\sim0.3-0.5$ dex and [$\alpha$/Fe]$=0.0-0.3$, it is possible to
account for the abundance difference of $0.5-0.8$ dex in the
transition region in Figure~\ref{gradient}.  At the high end (i.e.\
$0.8$ dex), this makes high inner disk metallicities consistent with
the very low abundances we derive.

\begin{deluxetable}{ccccccc}
\tablecaption{Metallicity gradient in NGC 7793. \label{alphafe_age}}
\tablewidth{0pt} 
\tablehead{ 
\colhead{Field} & \colhead{Age (Gyr)} & \colhead{[$\alpha$/Fe]} & \colhead{$a$} & \colhead{$\sigma_a$} & \colhead{$b$} & \colhead{$\sigma_b$}} 
\startdata 
SE & 8 & 0.0 &  0.004 & 0.011 & -1.49 & 0.10 \\
SE & 8 & 0.3 &  0.006 & 0.011 & -1.69 & 0.11 \\
SE & 12 & 0.0 & 0.005 & 0.012 & -1.64 & 0.11 \\
SE & 12 & 0.3 & 0.005 & 0.011 & -1.82 & 0.11 \\
NW & 8 & 0.0 &  0.095 & 0.023 & -2.17 & 0.20 \\
NW & 8 & 0.3 &  0.097 & 0.024 & -2.37 & 0.21 \\
NW & 12 & 0.0 & 0.100 & 0.023 & -2.35 & 0.21 \\
NW & 12 & 0.3 & 0.100 & 0.024 & -2.53 & 0.21
\enddata 
\tablecomments{Gradients are in the form of [$\alpha$/Fe]$=aR+b$,
where $a$ and the corresponding error have units of dex
kpc$^{-1}$. Abundance gradient for the inner disk using the data on
[O/H] abundances of HII regions from \citet{edmundspagel84} and
\citet{webstersmith83} is $-0.11\pm0.02$ dex kpc$^{-1}$}
\end{deluxetable}

Furthermore, in deriving metallicities of \HII regions,
\citet{edmundspagel84} and \citet{webstersmith83} use strong-line
abundance indicators.  These have been shown to result in
metallicities which are a factor of two ($0.3$ dex) higher than
nebular ($T_e$ based) or stellar abundances \citep{bresolin09b} and
likely enhance the difference between the inner (strong-line) and
outer disk (stellar) abundances in Figure~\ref{gradient}.

\begin{figure}[t]
\epsscale{1.1}
\plotone{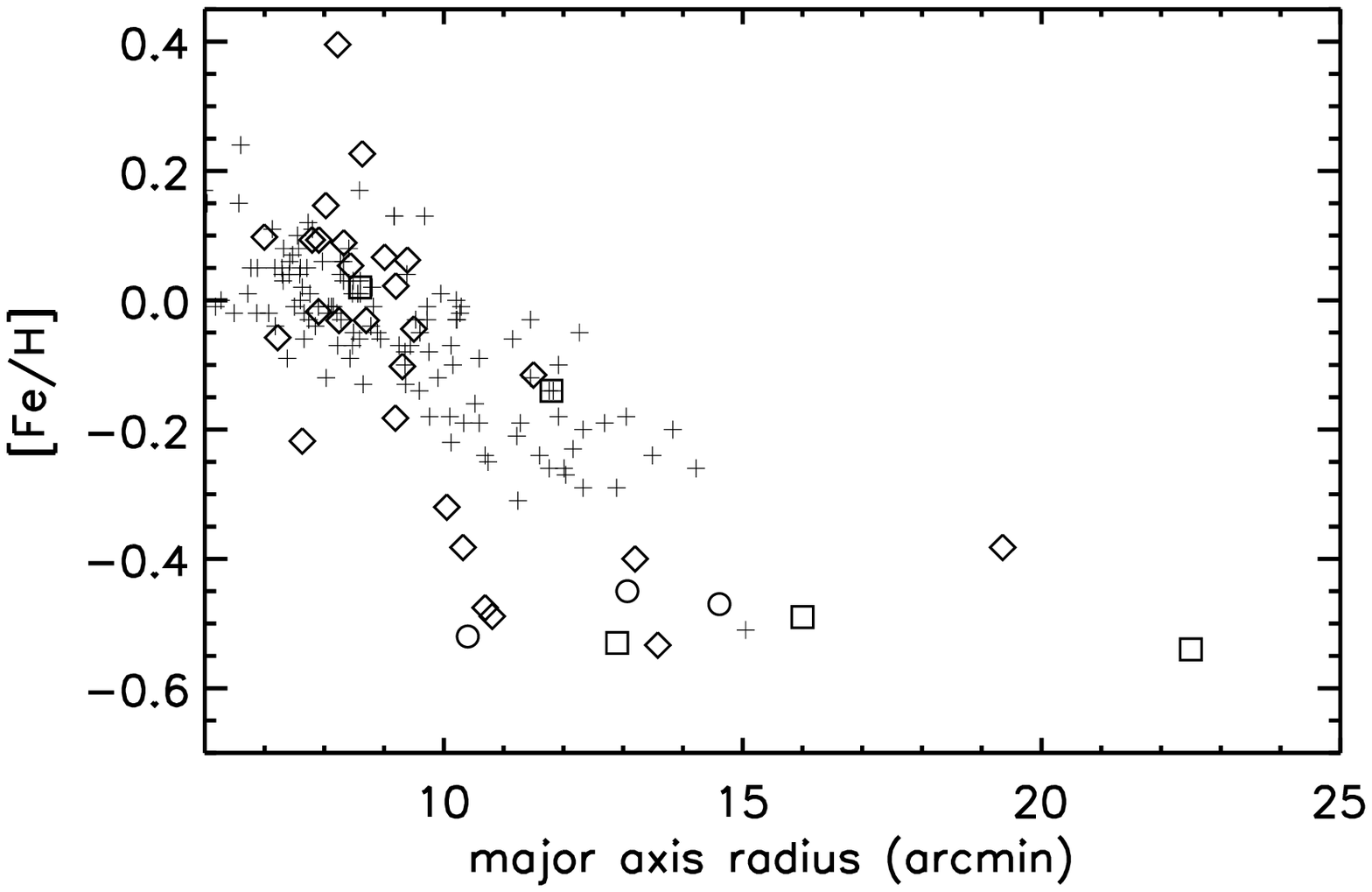}
\caption{Metallicity gradient in the disk of the Milky Way.  Plotted
are data for (i) old open clusters \citet[from][]{twarog97,chen03} as
diamonds and \citet{yong05} as squares, (ii) field red giants
\citep{carney05} as circles, and (iii) Cepheids from
\citet{andrievsky02b}, \citet{andrievsky02c}, \citet{luck03}, and
\citet{andrievsky04} as crosses.  The discrepancy is obvious between
old star abundances (diamonds, squares and crossed) and Cepheids
(crosses) in the region $10-15$ kpc. \label{mw}}
\end{figure}

Similar abundance behavior (although of a smaller magnitude) is
observed in the Milky Way.  In Figure~\ref{mw} we reproduce Figure~4
from \citet{carney05}.  Old open clusters are shown as diamonds
\citep[from][]{twarog97,chen03} and squares \citep[from][]{yong05}.
Circles are three field red giants from \citet{carney05}.  Cepheids
from the works of \citet{andrievsky02a}, \citet{andrievsky02b},
\citet{andrievsky02c}, \citet{luck03}, and \citet{andrievsky04} are
shown as crosses.  Cepheids follow the well-defined negative gradient
out to $\sim15$ kpc.  On the contrary, old stars exhibit steeper
gradient in the inner disk (out to $10-12$ kpc) and a flattening in
the outermost parts.  In the region $10-15$ kpc there is a clear
disconnect of up to $0.4$ dex between the Cepheid abundances and those
of old stars, similar to what we observe in NGC~7793.

\section{DISCUSSION \& CONCLUSIONS}
\label{sec:discussion}

We can summarize the results presented thus far as follows:

(i) The outer disk of NGC~7793 is primarily old with red giant branch
stars as the dominant stellar population and a small contribution from
asymptotic red giant branch stars.

(ii) After the contamination from faint background galaxies has been
taken into account, number counts of RGB stars, as well as those
derived using a matched catalog of all stars in the field, extend out
to $\sim10-11'$, or $\sim10.5-11.5$ kpc.

(iii) The effective surface brightness profile derived from star
counts traces the disk of NGC~7793 out to $9$ disk scale lengths and
is $\sim3$ mag arcsec$^{-2}$ deeper than the surface photometry data
of \citet{carignan85}.  Any potential break in the light profile may
be associated with a specific stellar population, but we see no
evidence of a truncation in old stars.

(iv) The metallicity gradient in the outer disk of NGC~7793 does not
exhibit a negative profile, characteristic of the inner galactic
regions.  The value of the slope is independent of the exact set of
stellar evolutionary tracks used (although the older, more
$\alpha$-enhanced isochrones result in lower overall metallicities).

(v) The outer disk metallicity gradient is in disagreement with the
inner disk slope.  The inner and outer disk abundances in the overlap
region are potentially in agreement after [$\alpha$/Fe] ratio,
age-metallicity relationship and the use of specific abundance
indicators have been taken into account.\\


Results presented in \S\ref{sec:mdf} suggest that the abundance
gradient derived from outer disk stellar [Fe/H] metallicities differs
in slope from the gradient calculated using [O/H] abundances from
inner disk HII regions (Figure~\ref{gradient}).  However, the two
cannot be directly compared as the latter dataset probes recent gas
abundances, while the former refers to chemical composition in stars
that are at least a couple of Gyr old.  In addition, given the results
of \citet{roskar08b}, which suggest that gas and stellar metallicities
are decoupled and follow opposite trends, it is difficult to conclude
whether our results point to an overall abundance gradient that gets
shallower or steeper with time.  \citep[Models of galactic chemical
evolution are successful in reproducing both trends,
e.g.][]{molla97,boissierprantzos99,portinarichiosi99,tosi88,chiappini01}
However, our results support the scenario presented by
\citet{roskar08b} in which gas abundances become steeper with time
(this is consistent with relatively steep inner disk slope in
NGC~7793) and the stellar abundance gradient in old stars is shallower
than that in young stars.


Although the stellar metallicity is the primary factor influencing
colors of RGB stars, age-metallicity degeneracy and the assumption of
single age in calculating the metallicity distribution function
introduce uncertainties in the derived MDF.  As shown in
Figure~\ref{gradient}, the derived metallicity gradient is practically
independent on adopted isochrones, assuming that age gradient over the
extent of the disk is close to constant; a non-zero age gradient would
result in a different metallicity profile.  Negative age gradient in
the outer disk of NGC~7793 would suggest that the real abundance
gradient has a higher slope than derived under a constant age
assumption.  Stellar ages which decrease with radius are indeed
consistent with the inside-out scenario for galaxy formation
\citep{larson76,matteuccifrancois89,chiappini97,naabostriker06,munozmateos07}.
In this picture, a galaxy's inner regions are built up at earlier
times than outer parts, and as a result contain on average older stars
than outermost regions.  However, recent results from resolved stars
\citep{barker07,williams09,williams10} and surface photometry
\citep{bakos08} seem to suggest that positive age gradients are
frequently observed in outer disks of spirals.  This is supported by
recent simulations of disk evolution
\citep{roskar08a,roskar08b,sanchezblazquez09}, which find that radial
migrations of stars within the disk are responsible for the reversed
age profile at large radii.  If the same holds in NGC~7793, the true
abundance gradient would be negative, flat or mildly positive,
depending on the magnitude of this effect.


There is a broad agreement that negative stellar abundance gradients,
easily explained in the context of inside-out models for galaxy
formation \citep{goetzkoeppen92,matteuccifrancois89}, are a common
feature of disk galaxies \citep{zaritsky94,ferguson98,gogarten10}.
Surface density, yield and star formation all decrease with radius,
resulting in metallicity distribution that is more metal-rich in
central parts and decreases progressively towards the outer disk.
However, abundance profiles in faint outer disks are more difficult to
derive and there is no general consensus on their shape and origin.
Growing body of evidence suggests that (most) spirals exhibit a
flattening of their metallicity gradient in the outermost disk.
Observationally, the strongest case has been made for the Galaxy
\citep{andrievsky04,yong06,carraro07,pedicelli09}, M83
\citep{bresolin09}, and M31 \citep{worthey05}.  In the models of
\citet{roskar08a}, \citet{roskar08b} and \citet{sanchezblazquez09},
stellar radial mixing has been shown to be able to produce flat
abundance profiles by 'smoothing out' the underlying negative
gradient.  On the other hand, a mildly positive metallicity gradient
has been observed in NGC~300 \citep{vlajic09}.

As mentioned earlier, a positive age gradient in the outer disk of
NGC~7793 would bias our derivation of abundance profile and a flat
underlying metallicity profile would be observed as a positive
gradient instead.  Our positive metallicity gradient in the NW field
could therefore be interpreted as a combination of a flat abundance
and positive age gradient.  This particular combination of age and
metallicity behavior has been found to arise as a consequence of
stellar migrations \citep{roskar08b,sanchezblazquez09}.  On the other
hand, it is possible that the positive metallicity gradient in
NGC~7793 is real and does not reflect the effects of age-metallicity
degeneracy.  \citet{minchev10b} find that the overlap of spiral and
bar resonances in the disk triggers significant migration of stars and
results in positive abundance profile in the outermost regions,
similar to what we observe in NGC~7793.


Alternatively, an external mechanism could be responsible for the
shape of the metallicity gradient in outer disks of spirals.  In
NGC~7793 in particular, the origin of a particular abundance profile
could be explained by the fact that the galaxy harbors a surprisingly
small \HI disk.  While a great majority of spirals have more or less
extended \HI disks, sometimes stretching out far beyond the known
optical edges, neutral hydrogen in NGC~7793 is detected only out to
$\sim11.5'$ \citep{carignanpuche90,walter08}, covering practically the
same radial extent as our stellar photometry.  (In addition, NGC~7793
exhibits a decreasing velocity curve in its outermost parts, which is
highly unusual for a galaxy of its size.)  The reason for a relatively
modest \HI disk in NGC~7793 is unclear, particularly given that the
galaxy has no obvious interactions that could have potentially
stripped the gas and truncated its \HI distribution.  Evidence for
stripped stars in the outer disk of NGC~7793 is also lacking.
However, it is possible to imagine that the upturn in the abundance
gradient in the NW field is a consequence of a dispersed stream of
stars that have long fallen below the detectability threshold in
surface brightness, but still pollute the outer disk metallicities.
Similar to stellar age, positive gradient in [$\alpha$/Fe] would --
under the assumption of constant [$\alpha$/Fe] -- be disguised as a
positive gradient in metallicity.  Accounting for the [$\alpha$/Fe]
increase from zero to $0.3$ over the range covered by our data would
likely not result in a flat abundance gradient, but would certainly
lower the slope of the gradient we derive for the NW field.


\begin{figure}[!t]
\epsscale{1.0}
\plotone{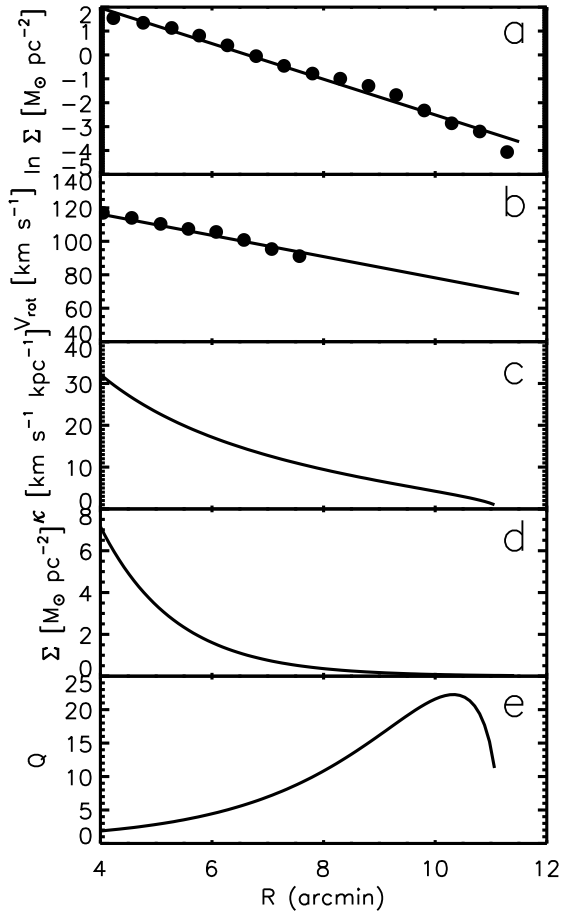}
\caption{(a) Logarithm of surface density in the outer disk of
NGC~7793.  Filled circles are data from \citet{puchecarignan90}; solid
line is a linear fit to these data points. (b) Same as (a), but for
rotational velocity. (c) Epicyclic frequency derived from the velocity
curve in (b). (d) Surface density. (e) Toomre $Q$
parameter. \label{toomre}}
\end{figure}

Finally, the shape of the metallicity gradient in the outer disk could
be primordial, originating in specific galaxy formation processes
taking place at high redshift.  \citet{cresci10} show that rotating
systems at z$\;\approx3$ show positive abundance gradients.  These are
presumed to be generated by cold streams depositing pristine material
into the centers of galaxies.  In today's galaxies the early positive
gradient in galactic center is reversed through processes of star
formation and chemical evolution, while the signs of the early
gradient remain in the outskirts of galaxies.


From the perspective of potential for star formation or radial
migration in a given region of a spiral disk, it is interesting to
examine the radial behavior of the Toomre $Q$ parameter.  In a thin
differentially rotating disk, rotation and pressure work to stabilize
the disk against axisymmetric perturbations.  On the other hand, the
disk is destabilized by its own surface density.  The disk is
considered unstable against axisymmetric modes if $Q$, given as:

\[ Q(r)=\frac{\sigma(r)\kappa(r)}{14.45\Sigma(r)} \]

\noindent is less than unity.  Here, $\sigma$ is radial velocity
dispersion in km s$^{-1}$, $\kappa$ is epicyclic frequency in the
units of km s$^{-1}$ kpc$^{-1}$, and $\Sigma$ is gas surface density
in M$_{\odot}$ pc$^{-2}$.  We employ \HI observations of
\citet{puchecarignan90} to estimate $\kappa(r)$ and $\Sigma(r)$.  We
approximate velocity curve and log of \HI surface density beyond $4'$
as linear functions of radius (Figure~\ref{toomre}a,b) and use these
fits to calculate $\kappa$ and $Q$.  Epicyclic frequency $\kappa$ is
calculated as:

\[ \kappa^2 = 2\Big(\frac{v^2}{r^2}+\frac{v}{r}\frac{{\mathrm d}v}{{\mathrm d}r}\Big) \]

\noindent and its radial distribution is shown in
Figure~\ref{toomre}c.  We estimate velocity dispersion in the outer
parts of NGC~7793 from the THINGS survey \citep{walter08} and adopt
$\sigma(r)=6$ km s$^{-1}$ as a mean velocity dispersion in our
observed field.  Radial distribution of the $Q$ parameter is shown in
Figure~\ref{toomre}e.  While using a declining velocity dispersion
(rather than constant as we do here) would result in a slower increase
of $Q$, Toomre parameter in the outer disk of NGC~7793 seems to be
significantly above unity and as such does not support an environment
in which there is an ongoing large scale star formation.  It is
possible however that the azimuthally averaged $Q$ profile does not
adequately describe star formation in outer disk.  At very low gas
densities in outskirts of spirals, star formation might have proceeded
in a stochastic manner and in small clumps, rather than on global
scales addressed by $Q$ as calculated here.  This is supported by
GALEX results of clumpy star formation in outer disks of nearby
spirals \citep[e.g.][]{thilker07}.  The high value of $Q$ presents a
potential problem for radial migration as well.  External
perturbations, e.g.\ from a passing satellite galaxy, have been shown
to be able to cause radial mixing in a Milky Way-type galaxy
\citep{quillen09} and could be responsible for stellar migrations in
high-$Q$ environments.

\acknowledgments We would like to thank the referee for detailed
comments which contributed to the quality of the paper.  Based on
observations obtained at the Gemini Observatory, which is operated by
the Association of Universities for Research in Astronomy, Inc., under
a cooperative agreement with the NSF on behalf of the Gemini
partnership: the National Science Foundation (United States), the
Science and Technology Facilities Council (United Kingdom), the
National Research Council (Canada), CONICYT (Chile), the Australian
Research Council (Australia), Minist\'erio da Ci\^encia e Tecnologia
(Brazil) and Ministerio de Ciencia, Tecnolog\'ia e Innovaci\'on
Productiva (Argentina)

\bibliography{marija}
\bibliographystyle{apj}

\end{document}